# Transformation of germanium to fluogermanates

[1]**S. Kalem\***, [2]**Ö. Arthursson**, [3]**I. Romandic**

[1]TUBITAK – UEKAE, The Scientific and Technical Research Council of Turkey – National Institute of Electronics and Cryptology, Gebze 41470 Kocaeli, TURKEY
[2]Department of Microtechnology and Nanosciences, Chalmers University of Technology, Göteborg, Sweden
[3]UMICORE, Electro-Optic Materials, 2250 Olen, Belgium

## ABSTRACT

The surface of a single crystal Germanium wafer was transformed to crystals of germanium fluorides and oxides upon exposure to a vapor of HF and $HNO_3$ chemical mixture. Structure analysis indicate that the transformation results in a germanate polycrystalline layer consisting of germanium oxide and ammonium fluogermanate with a preferential crystal growth orientation in <101> direction. Local vibrational mode analysis confirms the presence of N-H and Ge-F vibrational modes in addition to Ge-O stretching modes. Energy dispersive studies reveal the presence of hexagonal **α**-phase $GeO_2$ crystal clusters and ammonium fluogermanates around these clusters in addition to a surface oxide layer. Electronic band structure as probed by ellipsometry has been associated with the germanium oxide crystals and disorder induced band tailing effects at the interface of the germanate layer and the bulk Ge wafer. The acid vapor exposure causes Ge surface to emit a yellow photoluminescence at room temperature.

# 1     Introduction

Previous studies have shown that the surface of a crystalline Silicon could be transformed to a porous Silicon [1] or ammonium silicon hexafluoride $(NH_4)_2SiF_6$ by exposing Si wafer surface to a vapor of HF and $HNO_3$ chemical mixture [2-4]. This topic has been the focus of research interest for the unique optical and structural properties and for the possibility of application of the resulting material [2-4]. The presence of a cryptocrystalline structure soaking all the visible light, photoluminescence emission, self-assembly of low-dimensional structures and the possible applications as a dielectric insulation material in integrated circuits are among interesting features [2, 5]. It is of crucial importance to see if Germanium (Ge), another element of the same periodic group (IV) could be undergone similar surface modification under an acid vapor exposure consisting of HF and $HNO_3$. The result might provide fruitful insight into the understanding of the transformation dynamics and also offer interesting application routes for germanium. In this report, we show how the transformation of single crystalline p-type Ge wafer to germanates, that is ammonium hexafluorogermanate or ammonium germanium fluoride, $(NH_4)_2GeF_6$ (AGeF hereafter) and an hexagonal α-phase germanium oxide (α-$GeO_2$) could be realized by exposing Ge wafer surface to the vapors of HF:$HNO_3$ solution. The formation of crystalline germanates was confirmed by x-ray diffraction patterns and the vibrational modes of related species by Fourier transformed infrared (FTIR) analysis. The architecture of the crystalline structure at the surface and interface was determined by scanning electron microscopy (SEM). The incorporation of fluorine F, oxygen O and nitrogen N was also confirmed by Energy Dispersive Spectroscopy (EDS) measurements at SEM. Photoluminescence and ellipsometry analysis are also provided in order to determine the electronic band structure of the resulting material and the interface between the fluogermanate layer and Germanium bulk crystal.

## 2      Experimental

The formation of the germanates was realized in a Teflon cell from both n-type and p-type Ge (100) wafers using the vapor of a $HF:HNO_3:H_2O$ chemical mixture of semiconductor grade HF and $HNO_3$ with %48 and %65 by weight, respectively. The fundamentals of a vapor phase exposure process or vapor etching were based on our earlier work and the details of the related method and the Teflon container are described elsewhere [1]. No initial treatment is required for the Ge wafer but, just before adding the water, the $HF:HNO_3$ solution was primed for about 10 second using a small piece of p-Ge wafer. No electrical contacts or immersion to a liquid are required for this process.

## 3      Results and discussion

For the formation mechanism of the germanates under acid vapor treatment, the transformation can be described using the following overall chemical reaction by taking into account the final products, that is the fluogermanate $(NH_4)_2GeF_6$ and the germanium oxide $GeO_2$ as observed by XRD, EDS and FTIR measurements:

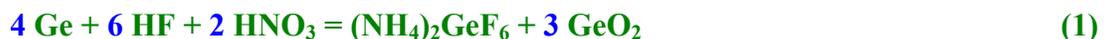

$$4\ Ge + 6\ HF + 2\ HNO_3 = (NH_4)_2GeF_6 + 3\ GeO_2 \qquad (1)$$

Concerning intermediate reactions leading to Eq.(1), there is no clear explanation due to the lack of experimental data for the difficulty of *in-situ* monitoring of surface chemical reactions. However, by analogy to previous works, one can suggest that the transformation process starts with oxidation of Germanium and the etching of the oxide by HF[5, 6]. Similar initial reaction pathways have been suggested by several research groups in chemical etching experiments of Silicon [6-10]. We think that the presence or condensation of $H_2O$ at the wafer surface should be carefully considered due to the solubility of $(NH_4)_2GeF_6$ in water. Actually, if one compares the vapor pressures, we find that the partial pressures of HF(130 kPa) and $HNO_3$ (8,3 kPa) are significantly higher than that of the $H_2O$ (3,2 kPa) [11], suggesting probably negligible involvement of $H_2O$ and plenty of HF in the vapor phase. Also, we should not rule out the presence of intermediate species such as $NH_4F$ which might be formed in the vapor phase following a reaction between HF and $HNO_3$ just prior to surface interaction. The $NH_4F$ molecules can subsequently interact with $GeO_2$ resulting in above mentioned fluogermanate and oxide clustering. In addition to this transformation, one can consider further effects such as an etching and removal of the surface germanate layer by HF, $HNO_3$ and even by $H_2O$ vapors.

Upon exposure to vapors of HF:$HNO_3$ (7:5) solution, the surface turns visible gold then hazy for an extended period of time. In this mixture, the ratio of HF and $HNO_3$ to the total amount of water $[H_2O]_T$ in the solution is $[HF]/[H_2O]_T = 0.62$ and $[HNO_3]/[H_2O]_T = 0.60$. The treatment of Ge surface by the acid vapor leads to the formation of germanates, namely ammonium germanium fluorides and oxides under various surface textures. The SEM images show the presence of some mesa-like semi-spherical surface structures or clusters on the surface (Fig.1a-c) and a formation of a layer of 5.4 μm, corresponding to a growth rate of around 230nm/minute. These mesa-like surface clusters have diameters of about 5 to 10μm and are composed of co-centric radial columnar structures as shown in the cross-sectional viewgraph (see Fig. 1b). A number of these clusters are grouped in the form islands on the surface, wherein multiple cracks are formed between the oxide clusters representing the presence of a significant strain field build-up between the layer and the bulk Germanium (Fig 1c). The remaining part of the layer is relatively smooth and consists of a germanium oxide.

Concerning the chemical nature of these clusters, energy dispersive analysis at SEM indicates they consist of only Ge and O atoms with an atomic concentration of oxygen up to 72%. The same analysis show that the ammonium fluogermanates are mainly located around these oxide clusters (Fig. 1c), suggesting that they are derived from Germanium oxide. The measurements indicate that these regions can contain up to 48 atomic % Fluorine and and 16 atomic % Nitrogen as evidenced from the EDS spectrum (Fig. 1e). The remaining surface area on Ge wafer consists of germanium oxide and germanium.

Another interesting feature of the surface is the formation of round shaped isolated regions of about 50μm as shown in Fig. 1(d) when the amount of $HNO_3$ is more than doubled (7:12:6). This chemical mixture indeed corresponds to a $HNO_3$-rich vapor wherein the chemical ratios are $[HF]/[H_2O]_T = 0.24$ and $[HNO_3]/[H_2O]_T = 0.56$. These areas consist of multiple crack lines which are stoped by the boundary of each region and are all parallel to each other as shown as an insert in Fig. 1(d). However, the same micrograph reveals that each particular region has a different crack line direction. Again, we observe the similar Ge oxide precipitation having cluster sizes of about 10μm, but this time they are evenly distributed over the surface.

Differential XRD pattern taken from the layer of germanate grown on p-type Ge wafer by vapor phase exposure is illustrated in Fig. 2. The layer reveals several major Bragg diffraction peaks with diffraction angle of 2θ at 25.5°, 35.6°, 37.6°, 39.0°, 41.4° and 48.2° which are indexed as (101), (111), (002), (201), (102) and (210) reflections, respectively. These peaks correspond to d-spacing of 3.495 Å, 2.521 Å, 2.393 Å, 2.305 Å, 2.179 Å and 1.886Å, respectively. By analogy to previous crystal data on powder fluogermanates[12] [13], the structure could be identified as the hexagonal $(NH_4)_2GeF_6$ crystal having a P-3m1 space group with crystal cell dimensions of a=5.51 Å, b=5.51 Å and c=4.52 Å. Note that these values of the cell parameters are about 6% lower than those reported for the powder form of $(NH_4)_2GeF_6$ which were prepared using $NH_4Cl$ and $GeO_2$ and HF [13]. The major peak is attributable to (101) reflection indicating the preferential growth direction of the fluogermanate polycrystalline layer. However, we find out that these patterns could well be the signature of an hexagonal α-phase quartz-like $GeO_2$ crystalline structure. The diffraction peaks observed in XRD spectrum of our sample are indeed very closely matching those of **α**-GeO$_2$ quartz-like crystal structure by 1% difference according to data provided in literature [13-16]. This suggests that there is a dominant contribution from the crystalline oxide clusters to the diffraction patterns. Moreover, we observe a broad background diffraction band between 60°-70° (see Fig.2) at around the peak of 2θ = 65.5° that is indexed as (203) for the reflection from the germanate layer. Dislocations, namely threading dislocations, small diameter crystals or strain could cause such a broadening of the XRD peaks. However, our Ge wafers as a whole are absolutely free from dislocations. Therefore, such dislocations, or strain might be induced after the processing. However, the broadening is just located around this region suggesting a different origin. In fact, several diffraction peaks of (NH$_4$)GeF$_6$ (61.7°, 63.4°, 66.3°) are also located in this region, proposing that the overall broadening was probably caused by an overlapping of peaks both from the oxide and the fluoride.

FTIR analysis of an AGeF layer which was prepared using the vapors of HF:HNO$_3$:H$_2$O (7:5) mixture reveals N-H, Ge-F and Ge-O related vibrations with features at 3240 cm$^{-1}$, 475 cm$^{-1}$, 577 cm$^{-1}$, 725 cm$^{-1}$, 833 cm$^{-1}$ and 1425 cm$^{-1}$ as shown in Fig. 3 and the possible band assignments are listed in Table-I. The strongest peak of the spectrum at 833cm-1 and its shoulders can be assigned to the asymmetric stretching modes of bridging oxygen Ge-O-Ge vibrations. The peak at 577cm$^{-1}$ and its shoulders were observed in hexagonal **α**-GeO$_2$ quartz crystals [16] but also Ge-F mode frequencies could contribute to this band of multiple peaks [17, 20, 21]. By analogy to previous FTIR studies on (NH$_4$)$_2$SiF$_6$ cryptocrystals and p-type Germanium, all the major features of the spectrum at 3240 cm$^{-1}$, 483 cm$^{-1}$, 725 cm$^{-1}$ and 1425 cm$^{-1}$ have been attributed to stretching, bending and rocking modes of N-H vibrations in NH$_4^+$ and GeF$_6^-$ ions [17-21]. There is no indication for the presence of Ge bonded to hydrogen inside the material as evidenced from the absence of Ge-H stretching modes in FTIR spectra. This result suggests that the interfacial layer between the germanates and Ge is most likely a porous germanium oxide layer. Using the relationship between the peak position of the Ge-O band and x as deduced for sputtered GeO$_x$[22], one can approximately estimate x to be between 1.3-2.6, indicating probably the presence of both an oxygen deficient and stoichiometric germanium oxide.

A relatively broad photoluminescence emission was observed at room temperature having a peak energy of around 560 nm (2.21 eV) with a half width at maximum of 125 nm (480 meV) as inserted in Fig. 4. Similar yellow photoluminescence emission has been observed in spark processed Ge [23] and stain-etched Ge [24]. These emissions were attributed to quantum confinement effects and to the combination of GeO$_x$ and Ge nanocrystals, respectively. We think that two possible sources can be responsible for this emission. First of all, some of the Ge clusters (probably covered by GeO$_2$) could be incorporated into the fluogermanate matrix during the transformation process, thus leading to such an emission. Secondly, a porous germanium oxide layer at the interface could be at the origin of this emission. The ellipsometry measurements indicate that the AGeF is rather transparent and the reflected signal is composed of the combination of effects involving both the bulk Ge and the oxide. The spectra in Fig. 4 show the important electronic critical points in interband transitions as deduced from the second derivative of the dielectric function for both the germanate layer and untreated p-type Ge. As indicated at the insert of this figure, there is relatively a good match (3 to 5%) between the principle gaps of the germanate layer and the bulk Ge. For the fluogermanate layer, the first direct gap, the spin-orbit splitted band and the X$_4$→X$_1$ direct transition gap are located at E$_1$=2.13 eV, E$_1$+ Δ$_1$ =2.38 eV (where Δ$_1$ is the splitting of 250meV, that is 8% higher than p-Ge) and E$_2$=4.56 eV, respectively. The critical point energies for the AGeF layer are few percent lower than those of the bulk Ge, thus indicating the absence of any quantum size effect. These lower band gaps are probably caused by disorder induced band tailing effects. Moreover, the band widths are larger particularly at the zone center **M$_0$(0,0,0)** related transitions **E$_o$' (Γ$_{25'}$→** $\Gamma_{15}$) at around 3.5eV. The same disorder could be held responsible for these discrepancies or peak width at around 3.5 eV as well as the increase in $\Delta_1$. Furthermore, the high energy peak **$E_2$** at 4.56 eV corresponds to the band gap energy of $GeO_2$ [25] confirming the effect of the Ge oxide at high energies. Thus, the surface modification of germanium through the formation of germanium oxide is likely responsible for this type of band structure modification. However, the origin of the photoluminescence emission could well be associated with Ge nanocrystals embedded in the quartz $GeO_2$ matrix.

## 4  Conclusion

  Treatment of a germanium surface by an acid vapor containing of HF and $HNO_3$ leads to the growth of a polycrystalline germanate layer consisting of an ammonium fluogermanate and **α**-phase $GeO_2$ crystals. Fluogermanates mainly grow around the germanium oxide clusters, suggesting that the oxide plays the role of a seed for their formation. $HNO_3$-rich vapor favors evenly distributed germanium oxide clustering and the formation of isolated crack regions. The electronic band structure of the germanate layer can be described by taking into account the effect of a germanium oxide and structural disorder at the surface of germanium. Principal band gaps are lower than those of the bulk Ge suggesting the significant effects of disorder on the band edges. The germanate layer emits a yellow photoluminescence at room temperature.

**Acknowledgement:**
One of us (S.Kalem) gratefully acknowledge the infrastructure research funding under MC2ACCESS project of the European FP6 programmes. Special thanks are due to Prof. U. Södervall, project manager of the infrastructure project.

**FIGURE and TABLE CAPTIONS**

**Figure 1.** Sanning electron micrograph of the germanate layer (#GE301) reveals the presence of sponge like semi-hemispherical clusters**(a)**, which are composed of co-centric radial columnar structure**(b)** and the islands of multiple clusters**(c)**. The height of this mesa structure is about 5 micron. The layer was grown using vapors of HF:HNO3(7:5) solution. The surface micrograph of a p-type Ge treated in a $HNO_3$-rich $HF:HNO_3:H_2O$ solution(7:12:6) is also shown**(d)**. The insert is the magnified image of the round-shaped crack regions. The EDS spectrum obtained from a germanate layer wherein the N and F atomic concentrations are 16.06% and 45.98%, respectively**(e)**.

**Figure 2.** X-ray diffraction pattern of the germanate layer grown on p-type Germanium using a vapor of HF:HNO3(7:5) mixture.

**Figure 3.** FTIR spectrum indicating N-H and Ge-F vibrational modes of $NH_4^+$ and $GeF_6^-$ ions in $(NH_4)_2GeF_6$. The layer (#GE301) was prepared using the vapor of a mixture of $HF:HNO_3$ (7:5). Germanium oxide can be identified by the presence of a strong absorption band at around 850 $cm^{-1}$ and 575 $cm^{-1}$.

**Figure 4.** The second derivative of the dielectric function indicating important electronic critical point energies in interband transitions for both the germanate layer and untreated p-type Ge. The inserted figure is the photoluminescence emission at room temperature from the germanate layer.

**Table 1** Peak frequencies of vibrational modes observed by FTIR in ammonium fluogermanate layer on Germanium wafer and their likely assignments based on References [17-21]. VS=very strong, S=strong, w=weak.

**Table 2** Electronic critical point energies of germanate layer as determined from spectroscopic ellipsometry measurements. The data is compared with those measured on blank p-Ge substrate.